\documentclass[multphys,vecphys]{svmult}

\usepackage{makeidx}     
\usepackage{graphicx}    
\usepackage{multicol}    

\makeindex             

\begin{document}

\title*{Molecular gas in spiral galaxies}
\author{Michele D. Thornley
\institute{Bucknell University, Department of Physics, Lewisburg, PA
  17837 (mthornle@bucknell.edu)}}
%
%
\maketitle

\section{Abstract:}

In this review, I highlight a number of recent surveys of molecular
gas in nearby spiral galaxies. Through such surveys, more complete
observations of the distribution and kinematics of molecular gas have
become available for galaxies with a wider range of properties (e.g.,
brightness, Hubble type, strength of spiral or bar structure). These
studies show the promise of both interferometers and single-dish
telescopes in advancing our general understanding of molecular gas in
spiral galaxies.  In particular, I highlight the contributions of the
recent BIMA Survey of Nearby Galaxies (SONG).

\section{Introduction}

Molecular gas is arguably one of the most important constituents to
consider in any study of galaxy evolution.  It is the raw material
from which stars form, and it responds strongly to dynamical
influences which can move large amounts of gas.  Spiral galaxies,
in particular, provide a variety of environments in which to study the
processing of molecular gas. For instance, galactic bars are likely
important in fueling central star formation by promoting the flow of
gas inward.  The star formation resulting from the influx of gas may
in turn promote the development of a bulge-like structure, or even the
destruction of the bar itself
(e.g.,\cite{fb93},\cite{nsh96},\cite{ath03}).  In addition, the
presence of a bar or spiral potential may influence the large-scale
pattern of gas, and therefore new star formation
(e.g.,\cite{ee90},\cite{tk91},\cite{cls03}).

The symmetry of the H$_2$ molecule effectively limits the direct
detection of molecular material to the small fraction of warm
molecular gas typically found in the vicinity of shocks or star
formation.  Due to its relatively high abundance and low effective
excitation, the CO molecule has become a standard proxy for H$_2$, and
its emission (particularly from the J=1-0 transition) has been
detected in many galaxies, in some cases out to very large redshift
(e.g., z=6.419,\cite{bcnc03}).  There is still significant debate
about the appropriate conversion factor between CO integrated
intensities and H$_2$ surface densities in different environments (see
presentations by M. Guelin, F. Walter, and A. Bolatto in this volume;
see also, e.g.,
\cite{mb88},\cite{wil95},\cite{ari96},\cite{dht01},\cite{blg02}).
However, there is no question that CO observations play a key role in
discovering the whereabouts and motions of molecular gas in galaxies
outside the Milky Way.

In the last decade, the improved sensitivity of millimeter-wave
facilities has enabled studies of the distribution and kinematics of
molecular gas in galaxies on smaller spatial scales, at larger radii,
and over smaller velocity shifts.  In particular, the advent of a
number of surveys has significantly advanced the study of molecular
gas in nearby galaxies since the previous Zermatt Symposium. We now
have a full accounting of the molecular gas in Local Group spirals,
as well as a self-consistent accounting of the detailed molecular gas
distributions in wider samples of Hubble type, bar contribution, and
star formation activity. For the purposes of this presentation, I
include techniques such as On The Fly (OTF) mapping (with single-dish
telescopes) and mosaicking (for interferometers) as ``survey''
techniques.

\section{Surveys of Local Group Spirals}

In the disks of our nearest spiral neighbors, M31 and M33, we can
resolve populations and kinematics of Galactic giant molecular cloud
(GMC)  
analogs.  Their proximity also requires on the order of 1000 pointings
for a typical millimeter telescope or interferometer to cover the
angular size subtended by the star-forming disk. Therefore, complete
studies of these individual galaxies can rightly be called surveys,
and they provide the means to study cloud populations in external
galaxies in an unbiased way.

\subsection{M31}

M31 is the nearest spiral to our own, offering an alternative view of
a large, bright spiral galaxy.  The large inclination of M31 motivates
the study of molecular gas at high angular resolution, in order to
separate arm and interarm regions and assess the properties and
development of spiral structure. Recent efforts to attain a complete
assessment of the molecular gas in M31 have seen a dramatic
improvement in spatial resolution, from the first complete CO survey
at 1.7 kpc resolution in 1993 (with the CfA 1.2m,\cite{dkit93}) to a
higher resolution (200 pc) survey of the southwestern half of the
galaxy (with the FCRAO 14m, \cite{loin96},\cite{loin99}) and a full
survey at better than 90pc resolution (with the IRAM
30m,\cite{nein01}) in the last few years.  We can now assess the
properties of the spiral structure and molecular cloud population in
M31, and compare it with those of the Milky Way (see presentation by
M. Guelin, this volume). Furthermore, the wealth of kinematic
information over a wide range of radii can be used to constrain galaxy
properties such as the shape and rotation speed of the bulge
\cite{ber01},\cite{bl02}.

\subsection{M33\label{m33}}

M33 offers a contrasting view of the properties of a nearby,
later-type spiral.  After a seminal study of the properties of
molecular gas in the nuclear region of M33 \cite{ws89},\cite{ws90},
it has taken a decade to expand on this important work.  The recently
completed 759-field survey covering the star-forming disk of M33 at 50
pc resolution (with BIMA, \cite{engar03}) is the first flux-limited
sample of distinguishable GMCs in a spiral galaxy, and creates a database
of 148 GMCs with which to study the properties of molecular clouds in
M33. In turn, it has enabled a higher resolution (20 pc) study of a
high-mass sample of giant molecular clouds in M33 \cite{rosol03}.
These studies confirm that M33 seems to be devoid of the most massive
GMCs seen in the Galaxy, and enable comparisons with other tracers
such as HI and H$\alpha$ which can be used to constrain GMC lifetimes,
a gas depletion timescale, and cloud formation mechanisms.

\section{Surveys outside the Local Group}

As we leave the Local Group, we return to the more traditional
definition of a survey, that of studying many different objects rather
than many pointings on a single object.  Through surveys, we are
rapidly gathering enough information on molecular gas in spiral
galaxies to begin making general statements directly from statistical
samples.

\subsection{Single-dish Surveys}

Much of our basic understanding of the global properties of gas
distributions in spirals come from single dish surveys.  Single dish
telescopes are able to observe a significant fraction of a galaxy in a
single beam, and do not suffer the uncertainties of missing
large-scale flux inherent in interferometer studies. The groundwork
for our understanding of the molecular gas content of spiral galaxies
was laid by earlier surveys
(e.g.,\cite{braine93},\cite{sage93},\cite{young95},to name a few),
which indicated that molecular gas can be detected over a range of
Hubble types, and tends to be more enhanced in the centers of galaxies
than in the outskirts. Further, we find that molecular gas and star
formation tracers are generally closely associated, though
interpretations of the form of this association vary (e.g.,
\cite{rowyoung96},\cite{kennI},\cite{kennII},\cite{wongbl02}).

Recent single-dish surveys (e.g., \cite{dumk01}, \cite{pag01},
\cite{nishi01a}, \cite{nishi01b}, \cite{bls03},\cite{hafstut03})
highlight the fact that virtually all single-dish millimeter
telescopes are actively engaged in surveys of a wider range of
molecular gas properties in spirals.  In recent years, these and other
studies have been used to improve our measures of molecular gas in
very late type spirals (Scd-Sm), construct higher resolution central
rotation curves than are available from HI, and show the promise of
mapping variations in molecular gas excitation and density through
observations of higher rotational transitions of CO.

Single-dish telescopes have also been used increasingly in a mapping
mode such as On-The-Fly (OTF) mapping.  Rather that measuring only the
central surface density or the variation of surface density along the
major or minor axis, these studies map the detailed variation of the
emission from various parts of the galaxy.  An excellent example is
the fully-sampled mapping of 5 nearby spiral galaxies with the NRAO
12m (now the UASO 12m), including IC 342 and M83
(\cite{crosthI},\cite{crosthII},\cite{crosthIII}).  These single dish
studies discern individual arm and interarm regions, and can trace the
relationship of molecular and atomic gas in detail.  These studies
show smooth connections between the molecular and atomic gas
components, which suggests that the gas phase may be determined by
local variations in turbulent pressure or dynamical conditions (e.g.,
\cite{mal88}, \cite{elm93},\cite{elmpar94}, \cite{sof95},
\cite{hon95}, \cite {hidsof02}).

\subsection{Interferometric Surveys}

Large single-dish telescopes such as the IRAM 30m and the Nobeyama 45m
can achieve moderate angular resolutions of
$\sim$10-20$^{\prime\prime}$, corresponding to $\sim$1-2 kpc at
d=20~Mpc.  Therefore, achieving kiloparsec or even sub-kiloparsec
resolution in a large number of nearby spirals requires the use of
interferometers. The observations possible with current millimeter
interferometers provide important information for planning future
surveys with the Combined Array for Research in Millimeter-wave
Astronomy (CARMA) and the Atacama Large Millimeter Array (ALMA).

\paragraph{Circumnuclear Molecular Gas}

With significantly higher resolution, interferometric studies are well
suited to study central molecular gas properties, and their
relationship with central activity.  The progress of one of the most
recent such surveys, The NUclei of GAlaxies project (NUGA), has been
reported at this meeting by S. Garcia-Burillo and L. Tacconi.  A study
with the NRO and OVRO interferometers of CO emission in 10 barred and
10 unbarred spirals \cite{saka1},\cite{saka2} reveals that molecular
gas is more concentrated in barred than unbarred spiral galaxies, thus
providing the first statistical evidence for bar-induced inflow in
transporting gas to the centers of galaxies.  However, this study
found no correlation between central gas concentration and type of
nuclear activity, which may indicate that the resolution of the survey
(4$^{\prime\prime}$) was not sufficient to assess separately
the properties of the circumnuclear gas.

Recent dissertation projects by S. Jogee \cite{jog99} and A. Baker
\cite{bak00}, as well as the OVRO MAIN survey (see description in
\cite{jog01},\cite{bak03}) complement these studies by targeting
particular aspects of the interaction of molecular gas with other
activities in the centers of galaxies.  Jogee compared central
molecular gas mass and central star formation activity, and showed
that information about gas kinematics as well as the size of the gas
reservoir was necessary to determine whether or not a central
starburst would form.  The Baker study of molecular gas in AGN with
broad H$\alpha$ line emission, as well as the OVRO MAIN multi-line
survey, are being used to study central kinematic and heating conditions,
and to test which processes are most influential in determining
central activity in galaxies.

The Nobeyama Millimeter Array (NMA) has been used recently to target
spirals in the Virgo cluster, using the advantage of a common distance
to minimize ambiguities when comparing structures between galaxies
(\cite{vicsI},\cite{vicsII},\cite{vicsIII}), thus being able to
compare directly the inferred surface mass densities and effects
of interactions on molecular gas distribution and motions.

\begin{figure}
\includegraphics[height=16cm]{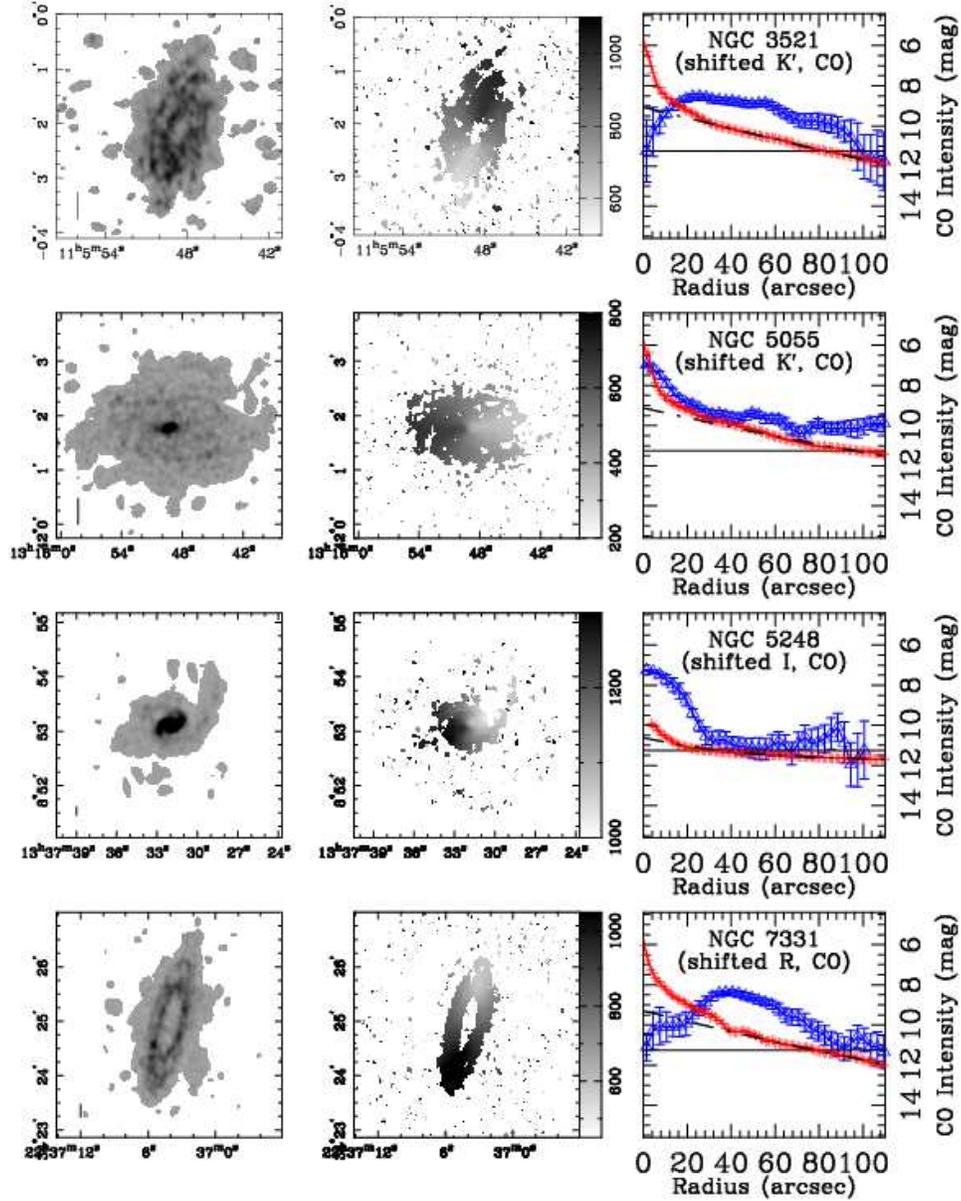}
\caption{BIMA SONG data for NGC3521, NGC5055, NGC 5248, and NGC7331
  (top to bottom). {\bf(left:)} integrated intensity (I$_{CO}$, with 1
  kpc bar at lower left); {\bf(center:)} velocity field (in km
  s$^{-1}$); and {\bf(right:)} azimuthally averaged radial profiles for
  I$_{CO}$ (triangles) and optical/NIR broadband light (crosses).  For
  the purposes of comparison, I$_{CO}$ has been converted to
  magnitudes for the radial profile (see \cite{reganI}) and the
  optical/NIR profiles have been shifted vertically. An exponential
  fit to the outer part of each optical/NIR profile has been
  overplotted as a dot-dash line.}
\end{figure}

\paragraph{Gas in Disks}

The interferometric surveys described thus far have included CO
brightness as a criterion for selection into the study.  Such studies
may be biased by an unknown factor toward galaxies with more
significant, or more highly concentrated, molecular gas masses.  The
BIMA Survey of Nearby Galaxies, or SONG, is the first large
interferometric CO survey to observe galaxies which were not chosen on
the basis of central CO brightness \cite{reganI},\cite{helfII}.  The
44 spirals in SONG were chosen by the following criteria: Hubble type
Sa-Sd, V$_{sys}<$ 2000 km s$^{-1}$, $i<70^o$, $\delta> -20^o$, and
B$_T<$11.0\footnote{SONG also imposed the criterion
D$_{25}<$70$^{\prime}$ to exclude M33, which is the subject of the
Engargiola et al. (2003) survey described in $\S$\ref{m33}.}.
Observations for SONG effectively mapped a $\sim$10-kpc diameter
region around each galaxy center, enabling a more accurate assessment
of the properties of gas in the inner disk.  CO J=1-0 emission was
detected in 41 of the 44 chosen galaxies (within a radius of
120$^{\prime\prime}$), providing a broad sample of galaxies with which
to explore the properties of gas and star formation over a range of
radii \footnote{The BIMA SONG data are available at the following
address:
http://nedwww.ipac.caltech.edu/level5/March02/SONG/SONG.html}.

The BIMA SONG data (see Figure 1) show that the molecular gas
distributions in spiral galaxies are very heterogeneous, and even the
azimuthally averaged radial profiles show a range of properties with
respect to the radial variation of the stellar light.  A comparison of
the molecular gas distributions in barred (SB/SAB) and unbarred (SA)
galaxies in the full SONG sample \cite{sheth04} confirms and
strengthens the results of the NMA-OVRO survey: the average central
gas surface density of barred spirals is three times higher than that
of unbarred galaxies.  It also appears that in many spirals the
increase in molecular gas surface density occurs over the same range
of radii in which the stellar light increases above that of an
exponential profile \cite{reganI},\cite{thorn02}.  As this central
``excess'' occurs commonly in unbarred galaxies as well as barred
galaxies, there may be some other mechanism than bar inflow required
to explain central gas excesses in all spiral types.  The complete
coverage of molecular gas in the inner disk has enabled detailed
studies of the relationship of molecular gas and star formation,
targeting both the validity and formulation of a star formation
``law'' (\cite{wong02}) as well as the influence of galactic bars and
gas inflow on the placement of star formation sites (\cite{sheth02}).
In addition, the wealth of kinematic information available has enabled
a systematic study of low-level streaming motions in a sub sample of
$\sim$20 SONG galaxies \cite{thorn03}.

\paragraph{Acknowledgements:}
I would like to thank my fellow members of the BIMA SONG consortium,
particularly T. Wong, M. Regan, and K. Sheth, for interactions that
continue to enhance the scientific value of SONG.  I would also like
to thank R. Magee and C. Spohn-Larkins, two Bucknell University
students who contributed to the research presented here.


\printindex

\end{document}